\documentclass[twocolumn,aps,pre,notitlepage,nofootinbib,longbibliography]{revtex4-2}

\usepackage[utf8]{inputenc}
\usepackage{graphicx}
\usepackage{natbib}
\usepackage{multirow}
\usepackage{amsmath,gensymb,amssymb}
\usepackage{siunitx}
\usepackage[bottom]{footmisc}
\usepackage{lipsum}
\usepackage{dcolumn}
\usepackage{xcolor}
\usepackage{upgreek}
\usepackage{float,ulem}

\usepackage{bm}
\usepackage{hyperref}
\hypersetup{
colorlinks=true, 
citecolor=magenta,
breaklinks=true, 
urlcolor= blue, 
linkcolor= blue, 
bookmarksopen=true, 
}
\usepackage{natbib}
\usepackage{color}
\usepackage[type={CC},modifier={by-nc-sa},version={4.0},]{doclicense}
\definecolor{linkcolor}{rgb}{0,0,0.6}

\begin{document}

\title{Experimental evidence of random shock-wave intermittency}	

\author{Guillaume Ricard}
\email{guillaume.ricard@u-paris.fr}
\affiliation{Universit\'e Paris Cité, CNRS, MSC, UMR 7057, F-75013 Paris, France}

\author{Eric Falcon}
\email{eric.falcon@u-paris.fr}
\affiliation{Universit\'e Paris Cité, CNRS, MSC, UMR 7057, F-75013 Paris, France}

\begin{abstract}
We report the experimental observation of intermittency in a regime dominated by random shock waves on the surface of a fluid. We achieved such a nondispersive surface-wave field using a magnetic fluid subjected to a high external magnetic field. We found that the small-scale intermittency of the wave-amplitude fluctuations is due to shock waves, leading to much more intense intermittency than previously reported in three-dimensional hydrodynamics turbulence or in wave turbulence. The statistical properties of intermittency are found to be in good agreement with the predictions of a Burgerslike intermittency model. Such experimental evidence of random shock-wave intermittency could lead to applications in various fields.
\end{abstract}

\maketitle

\textit{Introduction.---}
Intermittency is characterized by localized bursts of intense activity that even occur in relatively quiescent flows~\cite{Batchelor1949,Frisch1995}. It has been extensively investigated in the past decades, especially in three-dimensional (3D) hydrodynamics turbulence~\cite{Frisch1995} and has been ascribed to coherent structures such as vortex filaments~\cite{Batchelor1949}. Although describing successfully the energy cascade, the Kolmogorov dimensional analysis  \cite{Kolmogorov41} fails to explain the small-scale intermittency observed experimentally~\cite{Van1970,SinhuberPRL2017} and numerically~\cite{IyerPRL2021,Ishihara2009}. While several models have attempted to describe it~\cite{Kolmogorov62,Frisch1978,She1994}, the lack of closure of the Navier-Stokes equations lets the discussion widely open~\cite{Li2005,SinhuberPRL2017,IyerPRL2021}. Intermittency is a ubiquitous phenomenon that occurs in a wide range of experimental fields, e.g., wave turbulence~\cite{Falcon2007b,Falcon2022}, integrable turbulence \cite{RandouxPRL2014}, solar winds~\cite{Alexandrova2007}, Earth's magnetic field~\cite{DeMichelis2004} or atmospheric winds~\cite{MuzyPRL2010}, turbulent flames in combustion~\cite{Roy2023}, quantum turbulence~\cite{PolancoNature2021}, rotating turbulence~\cite{vanBokhovenPoF2009} or granular systems~\cite{Falcon2004}. 

Intermittency also occurs for Burgers turbulence, a simplified one-dimensional (1D) model of Navier-Stokes turbulence~\cite{Burgers1948}. Although less complex, Burgers turbulence is more predictable. It predicts the emergence of highly coherent structures, as random shocks governing its statistical properties, the energy spectrum, the probability distribution functions (PDFs) of velocity increments and gradients, as well as intermittency~\cite{Frisch2001}. In the inertial range, Burgers intermittency is predicted by a bifractal model and its origin is due to shock waves~\cite{Aurell1992,Gotoh1994,Bouchaud1995}. Numerical simulations of intermittency in the 1D stochastically forced Burgers equation have then been performed~\cite {Mitra2005,Buzzicotti2016,Murray2018}, but experimental evidence of intermittency in Burgers turbulence remains elusive so far, as a regime of random shock waves is hardly reachable experimentally.

Recently, we have experimentally shown that an ensemble of stochastic shock waves can emerge from random gravity-capillary waves on the surface of a fluid, made nondispersive using a magnetic fluid~\cite{Ricard2023}. Their fronts are not fully vertical, conversely to theoretical Burgers shock waves, and they drive the dynamics~\cite{Ricard2023}. {Here, we} explore the possible intermittent nature of such a random shock-wave-dominated field. We show that shock waves lead to intense small-scale intermittency that is quantified by the PDFs of the increment of the wave-amplitude fluctuations and by high-order statistics. In particular, the experimental structure-function exponents are found in quantitative agreement with a Burgerslike intermittency model, modified to take account of the finite steepness of the experimental shock-wave fronts. When the shock waves are removed by numerical post-processing, the nonlinear wave field is smoother and exhibits much weaker intermittency of a different nature. The latter is close to wave turbulence intermittency, reported experimentally~\cite{Falcon2007b,Falcon2022} and numerically~\cite{Meyrand2015}, involving other coherent structures (e.g., sharp-crested waves), and for which its origin is still a highly debated topic that may be related to the fractal dimension of these coherent structures~\cite{Connaughton2003,NazarenkoJFM2010}. Our experimental results appear thus of primary interest regarding the wide range of fields in which Burgers turbulence~\cite{Frisch2001} and wave turbulence~\cite{NazarenkoBook} occur. 

\textit{Theoretical background.---}
Intermittency corresponds to a continuous deformation over the scales of the PDFs of increments of a given field (e.g., fluid-velocity or wave-amplitude fluctuations)~\cite{Frisch1995}. Phenomenological models have been developed since the 60s to quantify such small-scale intermittency~\cite{Kolmogorov62,Frisch1995}. To {do so}, first-order increments of a temporal signal $\eta(t)$ are defined as $\delta\eta(t,\tau)=\eta(t+\tau)-\eta(t)$, where $\tau$ is the time lag. The scaling properties of the corresponding structure functions of order $p$ as $\mathcal{S}_p(\tau)\equiv\langle|\delta\eta|^p\rangle_t\sim\tau^{\zeta_p}$ (with $p$ a positive integer) are the keys quantities to characterize intermittency. A nonlinear dependence on the exponents $\zeta_p$ with $p$ is indeed a signature of intermittency. In 3D hydrodynamics, experiments showed a nonlinear scaling of $\zeta_p$ with $p$~\cite{Batchelor1949,Van1970} that is not described by the Kolmogorov nonintermittent prediction in $\zeta_p=\frac{n-1}{2}p$~\cite{Kolmogorov41}. Phenomenological models have been proposed to tackle this discrepancy and predict the nonlinear shape of $\zeta_p$~\cite{Kolmogorov62,Frisch1995,She1994,Frisch1978}. Some agree with experiments~\cite{Van1970}, but do not account for the observed oscillating $\zeta_p$~\cite{SinhuberPRL2017,IyerPRL2021}. For Burgers turbulence, the presence of shock waves leads to strong intermittency. In the limit of vanishing viscosity of the Burgers equation, the structure functions are then predicted by a bifractal model, since regular random waves and shock waves coexist, as~\cite{Aurell1992,Gotoh1994,Bouchaud1995}
\begin{align}    
& \mathcal{S}^{\mathrm{ni}}_p(\tau) \sim \tau^{(n-1)p/2} \ \ \mathrm{for} \ p<2/(n-1)\ , \label{Sp_Burg_eq} \\   
& \mathcal{S}^{\mathrm{B}}_p(\tau) \sim \tau^1 \ \ \mathrm{otherwise.} \label{Sp_Burg_eq2}
\end{align}
The scaling in $\zeta_p=(n-1)p/2$ in Eq.~\eqref{Sp_Burg_eq} comes from the smooth random component of the solutions~\cite{Frisch1995,Mitra2005}. It is also predictable by dimensional analysis and is related to the exponent $n$ of the energy spectrum. The second scaling in $\zeta_p=1$, regardless of $p>2/(n-1)$, occurs when the shocks dominate, and the unit value is related to the vertical feature of the shock fronts leading to $\delta$-Dirac singularities for its increments (see Appendix~\ref{BurgersNum}). Note that the effects of the spectral bandwidth of the random forcing to Burgers equation and of dissipation on the scaling of Eq.~\eqref{Sp_Burg_eq2} have been numerically investigated~\cite{Mitra2005,Verma2000}. Nevertheless, to our knowledge, no experimental evidence for such intermittency in Burgers turbulence has been established so far. 

\begin{figure}[t!]
    \centering
    \includegraphics[width=1\linewidth]{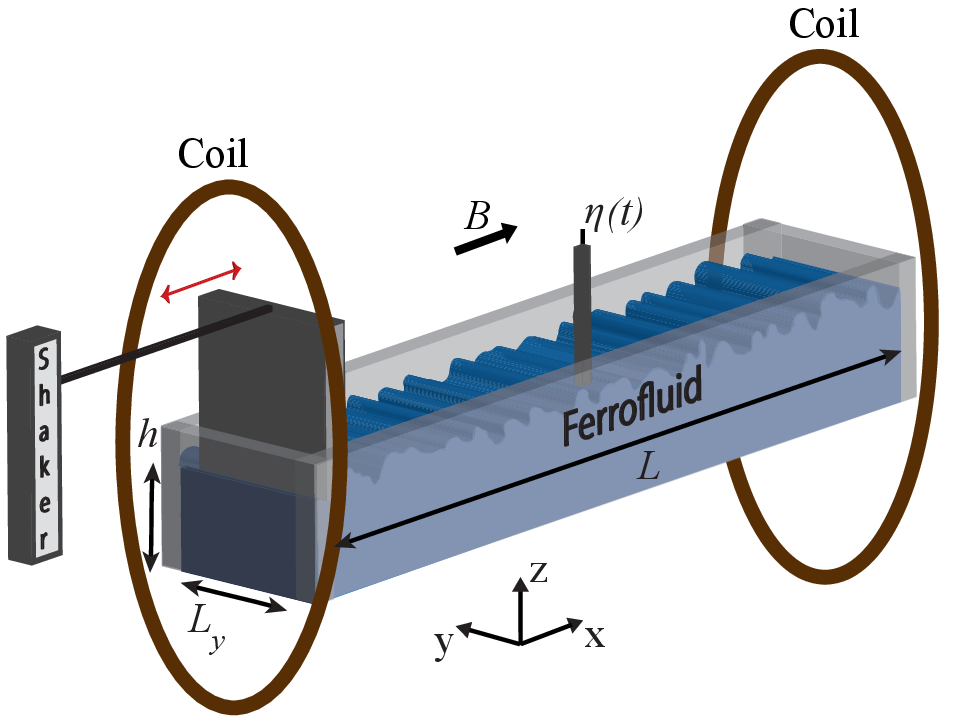} 
    \caption{{Experimental setup. A pair of Helmholtz coils generates a horizontal homogeneous magnetic field
B on the ferrofluid surface. Random waves are driven by a wave maker linked to a shaker at one end of the
canal. The wave elevation $\eta(t)$ is measured at a single point using a capacitive wire gauge. $L=15$ cm, $L_y=2$ cm, and $h=2$ cm.}}
    \label{setup}
\end{figure}

\textit{Experimental setup.---}
The experimental setup {is represented in Fig.~\ref{setup} and} has been described in detail in Ref.~\cite{Ricard2023}. A canal of length $L=15$~cm is filled with a liquid up to a depth of {$h=2$}~cm. An electromagnetic shaker with a paddle is located at one end to inject energy in a narrow random frequency bandwidth $f_0\pm\Delta F$, with $f_0=8.5$~Hz and $\Delta F=2.5$~Hz. {The typical rms wave maker amplitude is few mm.} We use a magnetic fluid (Ferrotec PBG400 ferrofluid) subjected to an external horizontal magnetic field {$B$ collinear to the wave propagation. We have previously shown that increasing the strength of the magnetic field enables us to tune the surface-wave dispersion relation $\omega(k)$, and to achieve, at large $B$, a nondispersive acousticlike regime in $\omega(k)\sim k$, where random shock waves dominate~\cite{Ricard2023}. This regime occurs when the dispersive gravity and capillary terms in the theoretical dispersion relation of surface waves are much smaller than the nondispersive magnetic term. For our range of experimental parameters and $B=650$ G, we showed that the magnetic term is about 20 times larger than the gravity and capillary terms, but weak dispersive effects are still present due to some residual gravity-capillary waves~\cite{Ricard2023}. Such a nondispersive wavefield regime has been also experimentally evidenced by computing the spatiotemporal spectrum of the wave elevation leading to a spectrum in agreement with $\omega(k)\sim k$ for $B=650$ G whereas the gravity-capillary dispersive relation is observed for $B=0$ G~\cite{Ricard2023}. 

Here, we keep constant the horizontal magnetic field strength to $B=650$ G to observe and characterize intermittency of the nondispersive regime involving mainly random shock waves.} We measure the surface elevation, $\eta(t)$, at a single point using a homemade capacitive wire gauge (0.2 mm in diameter, 10 $\upmu$m in vertical resolution at 2~kHz), {located in the middle of the canal,} during $\mathcal{T}=15$~min. {We checked that the location of the probe in the canal does not change the results reported here in particular for shock waves that conserve their shape (i.e., their discontinuity) along the canal {(see Appendix~\ref{shock_pulse_fomation})}.} To quantify the wave nonlinearity, we measure the {mean} wave steepness as $\epsilon\equiv\eta_{\mathrm{rms}} k_{m}$, with $\eta_{\mathrm{rms}}=\sqrt{\langle\eta(t)^2\rangle_t}$ the standard deviation of the surface elevation, and $k_{m}$ the wave number at the wave spectrum maximum (typically at the forcing scale)~\cite{Falcon2022}. $\epsilon$ is chosen constant to a low value of $0.07$ to be in a weakly nonlinear regime.

\begin{figure}[t!]
    \centering
    \includegraphics[width=1\linewidth]{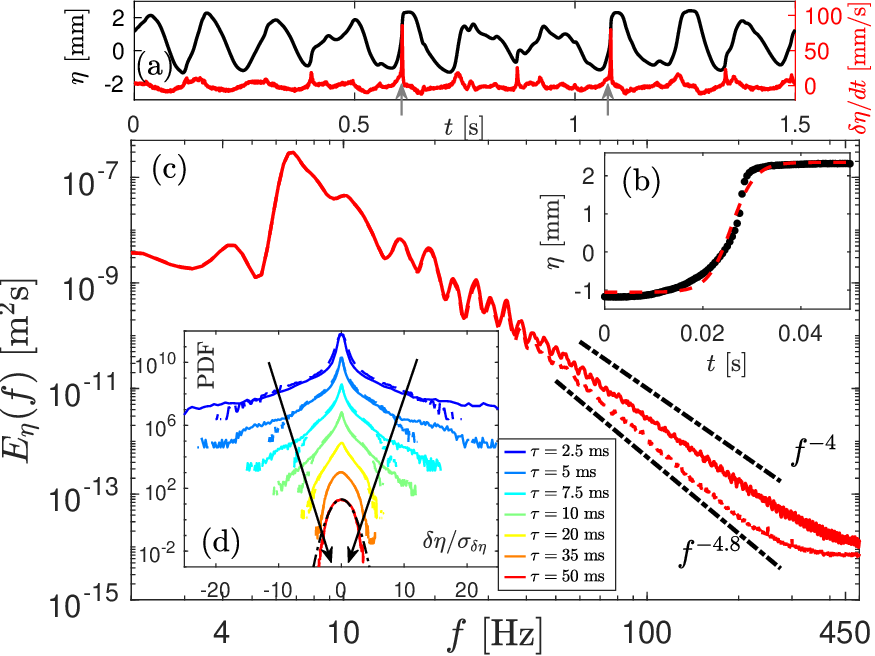} 
    \caption{(a) Typical signal $\eta(t)$ (black) and its difference (red) highlighting shock waves (arrows). (b) Zoom on a shock wave ($\bullet$) fitted by a Burgers solution in $a\tanh(t/t^*)$ ($a=1.7$~mm, $t^*=4.7$~ms). (c) Frequency-power spectra $E_\eta(f)$ with (solid line) and without (dashed line) shock waves. Black dash-dotted line{s}: Theoretical spectrum in $f^{-4}$ for second-order singularities~\cite{Ricard2023,KuznetsovJETP2004} {and best fit in $f^{-4.8}$ for the spectrum without shock wave}. (d): PDFs of the fourth-order increments $\delta\eta(\tau)$ for time lags $\tau\in[2.5, 50]$~ms (see arrows), i.e., $f\in[10,200]$~Hz, with (solid lines) and without (dashed lines) shock waves. Black dash-dotted line: Gaussian with unit standard deviation. Correlation time is $\tau_c\approx33$~ms (see Appendix~\ref{corr_time}). PDFs have been shifted vertically for clarity.} 
\label{Spectra}
\end{figure}

\textit{Wave energy spectra.---}
A typical example of the fluctuations of the surface elevation, $\eta(t)$, over time, is plotted in Fig.~\ref{Spectra}(a). Very steep wave fronts emerge as random shock waves (see arrows) corresponding to peaks in the signal difference. A typical shock wave is enlarged in Fig.~\ref{Spectra}(b) well fitted by a solution of the Burgers equation~\cite{Burgers1948} in $a\tanh(t/t^*)$ with $a$ its amplitude and $t$ quantifying its steepness. {These shock-wave parameters are inferred from a 15-min signal in which 863 shock waves are detected. We find $a=1.3\pm0.4$~mm and $t^*=5.7\pm2.1$~ms and Gaussian distributions for $a$ and $t^*$}. This very steep profile does not reach a fully vertical front (i.e., $t^*\rightarrow0$) that occurs only for vanishing viscosity in the Burgers equation. We have shown previously that shock waves are characterized by a discontinuity leading to a $\delta$-Dirac singularity in the second-order difference of their amplitude~\cite{Ricard2023}. Using the Fourier transform $\widehat{\eta}(\omega)$ of $\eta(t)$, the frequency-power spectrum, $E_\eta(\omega)\equiv |\widehat{\eta}(\omega)|^2/\mathcal{T}$, of the wave-amplitude fluctuations is computed and shown in Fig.~\ref{Spectra}(c). A well-defined power-law scaling in $E(\omega) \sim \omega^{-4}$ is observed. Such a spectrum of shock waves has been shown to agree with the theoretical Kuznetsov spectrum of second-order singularities~\cite{Ricard2023,KuznetsovJETP2004}. It differs from the classical  $\omega^{-2}$ Burgers spectrum {(see Appendix~\ref{BurgersNum})}, as the experimental shock-wave fronts are not fully vertical. When the singularities are removed from the signal by smoothing numerically the signal around the discontinuities {(using a moving-average filter)}~\cite{Ricard2023}, the corresponding spectrum scales in $\omega^{-4.8}$ [see red-dashed line {for data, and black dash-dotted line for the best fit,} in Fig.~\ref{Spectra}{(c)}] as it is mainly governed by residual gravity-capillary waves and dissipation. Let us now focus on the statistics of the increments of the surface elevation.

\textit{Probability density functions.---}
For such a steep $\omega^{-4}$ spectrum, high-order difference statistics are required to test intermittency~\cite{Falcon2010b}. We will use afterward the fourth-order increment $\delta^{(4)}\eta=\eta(t+2\tau)-4\eta(t+\tau)+6\eta(t)-4\eta(t-\tau)+\eta(t-2\tau)$. As we have checked, this is more than enough to achieve convergence of the structure functions, which then no longer depend on the higher order of the increment used~\cite{Falcon2010b}. $\delta^{(4)}\eta$ will be denoted $\delta\eta$ afterward, for the sake of clarity. The PDFs of $\delta\eta(\tau)$ are displayed in Fig.~\ref{Spectra}(d) for different time lags $\tau$ corresponding to more than one decade in frequency $f=1/(2\tau)$. The PDFs are found to be almost Gaussian at large $\tau$, as expected, and display a continuous deformation with decreasing $\tau$ leading to heavy tails at small $\tau$, as a signature of intermittency~\cite{Frisch1995}. Its origin is ascribed to the coherent structures, i.e., shock waves, storing energy, and traveling over the canal. Indeed, at small scales ($\tau<20$~ms, i.e., $f>25$~Hz) the PDF tails are heavier in the presence of shock waves (solid lines) than without shock waves (dashed lines). Intermittency appears thus much more pronounced in the presence of shock waves than without, as it will be quantified below by the structure-function analysis. Note that, although the shock waves are removed, heavy tail PDFs still remain since other (less intense) coherent structures are present (see below).
\begin{figure}[t!]
    \centering
    \includegraphics[width=1\linewidth]{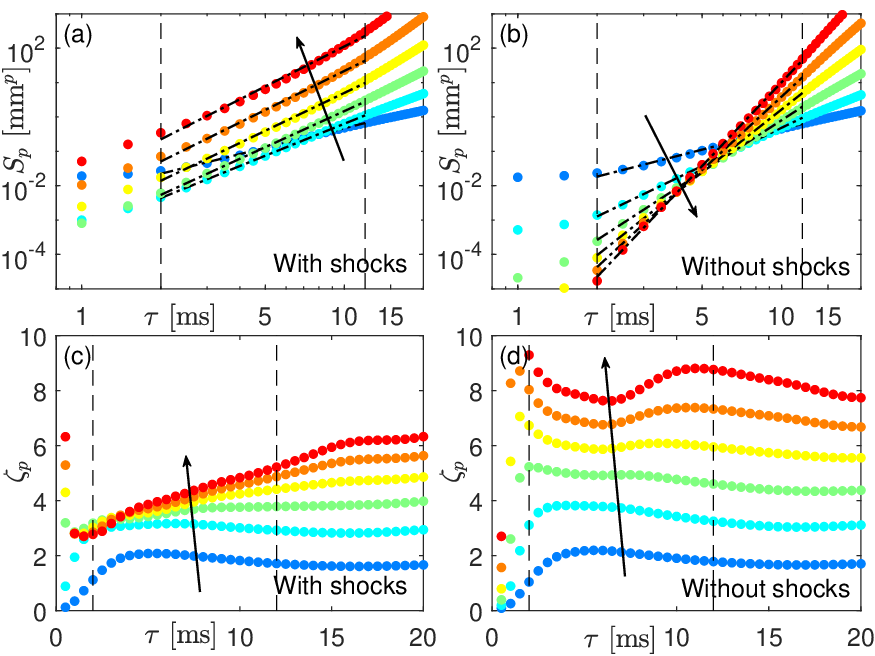} 
    \caption{Structure functions $\mathcal{S}_p(\tau)$ as a function of $\tau$ with (a) and without (b) shock waves, for increasing $p\in[1,6]$ (see arrow). Dash-dotted lines: Power-law fits in $\tau^{\zeta_p}$ within the inertial range (between vertical dashed lines). Exponent $\zeta_p$ versus $\tau$ with (c) and without (d) shock waves, for increasing $p\in[1,6]$ (see arrow).}
    \label{Sp_tau}
\end{figure}

\textit{Structure functions.---}
We now compute the structure functions $\mathcal{S}_p(\tau)$ in order to quantify the above-reported intermittency. $\mathcal{S}_p(\tau)$ are shown in Fig.~\ref{Sp_tau} with (a) and without (b) shock waves for $p\in[1,6]$. At first glance, power laws in $\tau^{\zeta_p}$ are observed in the inertial range. To extract more accurately these exponents, we compute logarithmic derivatives $\zeta_p(\tau)\equiv d\log(S_p)/d\log(\tau)$, i.e., the logarithmic local slopes, that are shown in Fig.~\ref{Sp_tau} with (c) and without (d) shock waves.

\begin{figure}[t!]
    \centering
    \includegraphics[width=1\linewidth]{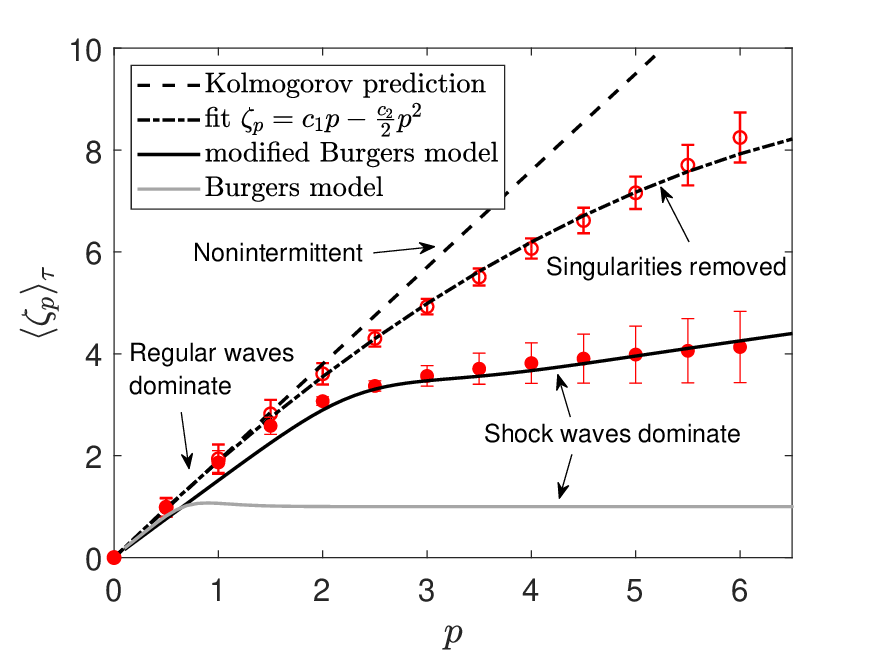} 
    \caption{Exponents $\langle \zeta_p \rangle$ of the structure functions versus $p$ with ($\bullet$) and without ($\circ$) shock waves. Gray line: Burgers model of Eqs.~\eqref{Sp_Burg_eq}-\eqref{Sp_Burg_eq2}. Solid line: modified Burgers model (see text). Dash-dotted line: nonlinear fit in $\zeta_p=c_1 p-\frac{c_2}{2} p^2$ with $c_1=2$ and $c_2=0.23$. Dashed line: Kolmogorov nonintermittent prediction of Eq.~\eqref{Sp_Burg_eq}. The errorbars correspond to the $\zeta_p$ standard deviation in the inertial range.} 
    \label{Zetap}
\end{figure}

With shock waves, a clear increase of $\zeta_p$ with $\tau$ is observed in the inertial range for $p>3$, whereas $\zeta_p$ is found to be roughly constant without shocks, as for classical intermittency~\cite{Frisch1995}. The values of the exponents averaged over $\tau$ in the inertial range, $\langle\zeta_p\rangle$, are reported in Fig.~\ref{Zetap}. In the presence of shock waves (full symbols), we find that $\langle\zeta_p\rangle$ strongly increases until $p\approx 2$ and much less for larger $p$. Note that this observed intermittency is much stronger than for 3D hydrodynamic intermittency~\cite{Van1970,SinhuberPRL2017,Ishihara2009,IyerPRL2021} or wave turbulence intermittency~\cite{Falcon2007b,Falcon2022}, as stronger nonlinearities are involved, as shock waves. Such an evolution of $\langle\zeta_p\rangle$ with $p$ could be qualitatively described by the bifractal model of Burgers turbulence of Eqs.~\eqref{Sp_Burg_eq}-\eqref{Sp_Burg_eq2}. Indeed, for low $p$, Eq.~\eqref{Sp_Burg_eq} predicts $\zeta_p=\frac{n-1}{2}p$ with $n$ the frequency power-law exponent of the energy spectrum. This nonintermittent scaling, {\it à la} Kolmogorov, comes from the smooth component of $\eta(t)$. For large $p$, Eq.~\eqref{Sp_Burg_eq2} predicts that $\zeta_p$ is independent of $p$, such intermittent scaling being ascribed to the shock waves~\cite{Aurell1992,Gotoh1994,Bouchaud1995}. However, a clear departure is observed between the experiments and the Burgers model of Eq.~\eqref{Sp_Burg_eq2} (see full symbols and gray line in Fig.~\ref{Zetap}). This discrepancy is due to the fact that the experimental shock-wave fronts are not fully vertical, which can be taken into account by modifying the Burgers model (see below). When the shock waves are removed, the experimental evolution of $\langle\zeta_p\rangle$ is strongly different (see empty symbols in Fig.~\ref{Zetap}). Burgers intermittency has vanished, but intermittency is still weakly present. Indeed, $\zeta_p$ can be fitted by the simplest nonlinear law, $\zeta_p=c_1 p-\frac{c_2}{2} p^2$, with $c_1=2$ and $c_2=0.23$~(see dashed-dotted line). $c_1$ is related to the Kolmogorov nonintermittent prediction $c_1\simeq(n-1)/2$ with $n=4.8$ (see dashed line). The nonzero value of $c_2$, quantifying the deviation from linearity, shows that much weaker intermittency remains as other coherent structures (i.e., steep gravity-capillary waves) are still present, as routinely reported in wave turbulence~\cite{Falcon2007b}. This is also confirmed by comparing the magnitudes of the structure functions with and without shock waves that differ from a few orders of magnitude at small $\tau$ once $p>3$ [compare Fig.~\ref{Sp_tau}(a) and \ref{Sp_tau}(b)]. The reason is that shock waves through their discontinuities, generate peaks in the increment of amplitude, $|\delta\eta|^p$, increasing with $p$ that thus dominate the $\mathcal{S}_p$ values. Another way to quantify intermittency, i.e., the PDF shape deformations over the scales, is to compute the coefficients of flatness, $\mathcal{S}_4/\mathcal{S}^2_2$, and hyperflatness, $\mathcal{S}_6/\mathcal{S}^3_2$. We find that they both confirm that strong intermittency is well ascribed to shock waves (see Appendix~\ref{Flatness}).

\begin{figure}[t!]
    \centering
    \includegraphics[width=1\linewidth]{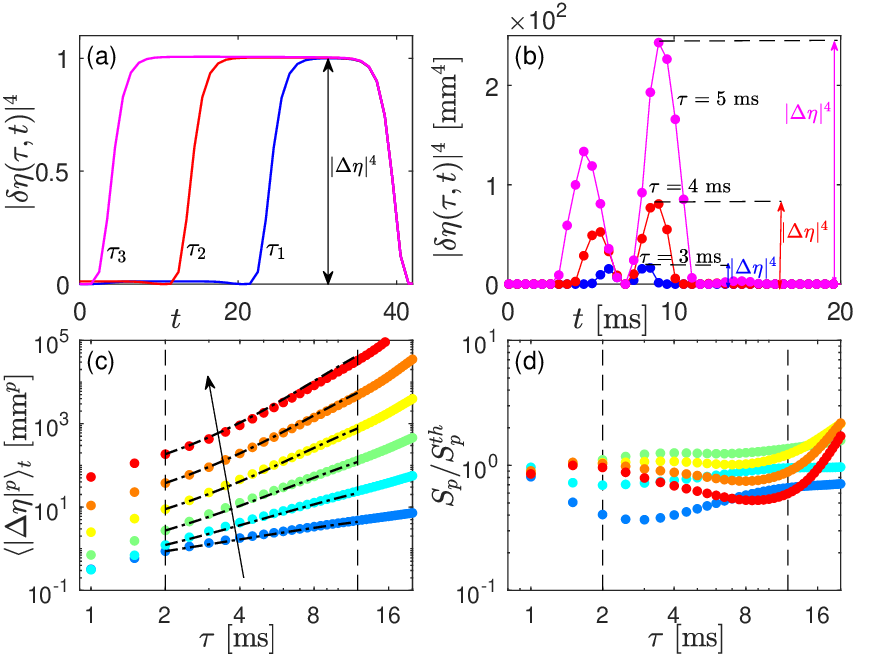} 
    \caption{Evolution of a single shock-wave increment amplitude for $p=4$, $|\delta\eta(t,\tau)|^4$ and different $\tau$: (a) simulations of the Burgers equation and (b) experimental data. (c) Experimental increment maximum amplitude, $\langle|\Delta\eta|^p\rangle_t$, versus $\tau$ for increasing $p\in[1,6]$ (arrow). Dash-dotted lines: predictions of Eq.~\eqref{Spthamp} {for a typical} shock-wave geometry of {$a=3.1$~mm} and {$t^*=7.7$~ms}. (d) Compensated structure functions $\mathcal{S}_p/\mathcal{S}_p^{\mathrm{th}}$ for $p\in[1,6]$. Vertical dashed lines indicate the inertial range.}
    \label{S4_SpNorma}
\end{figure}

\textit{Modified Burgers model.---}
The experimental behavior of $\zeta_p$ in Fig.~\ref{Zetap} deviates from the Burgers intermittency model~\cite{Aurell1992,Gotoh1994,Bouchaud1995}. This model assumes shock waves with vertical fronts, and the structure functions are then predicted, in the inertial range, by~\cite{Gotoh1994}
\begin{equation}
    \mathcal{S}_p^{\mathrm{B}}(\tau)=\Gamma \tau^1{\langle|\Delta\eta|^p\rangle}_{t}, \ \ \mathrm{for}\ \ p>2/(n-1),
    \label{SclaingBurgers}
\end{equation}   
where $\Gamma$ is the shock rate (mean number of shocks per second) and $|\Delta\eta|^p=|\delta\eta(\tau,t_s)|^p$ is the $p$th moment of the increment amplitude of a shock wave occurring at $t=t_s$. Equation~\eqref{SclaingBurgers} is only valid in the inertial range which is verified experimentally since the shock-wave singularity duration $\tau_s \sim 1$ ms $\ll \tau \ll$ duration between shocks $1/\Gamma \sim 1$~s. The shock-wave number and amplitude must be also large enough, as verified experimentally~\cite{Ricard2023}. We first solve numerically the 1D viscous Burgers equation~\cite{Burgers1948}. Burgers shock waves involve indeed vertical fronts (see Appendix~\ref{BurgersNum}), its increment amplitude is well independent of $\tau$ and its increment width scales as $\tau^1$ [see Fig.~\ref{S4_SpNorma}(a)], as predicted by Eq.~\eqref{SclaingBurgers}. {Note that the three nondimensional values of $\tau_i$ used in Fig.~\ref{S4_SpNorma}(a) are chosen to correspond to frequencies within the inertial range of the theoretical $\omega^{-2}$ Burgers power spectrum (see Appendix~\ref{BurgersNum}).}

Experimentally, the shock-wave fronts have a finite steepness. As a consequence, the increment maximum amplitude $|\Delta\eta|^p$ will depend on $\tau$, conversely to Eq.~\eqref{SclaingBurgers}. Indeed, Fig.~\ref{S4_SpNorma}(b) shows the experimental evolution of the increment amplitude for $p=4$, $|\delta\eta(\tau)|^4$, of a single shock wave, over time and for different $\tau$. We find that the widening of the shock width scales as $\tau^1$ as for the Burgers model, but a clear increase of its maximum amplitude occurs with $\tau$, contrary to the Burgers case [Fig.~\ref{S4_SpNorma}(a)]. Assuming that shock waves are well described by a Burgers solution in $\eta(t)=a\tanh(t/t^*)$ with $t^*$ its typical steepness [see Fig.~\ref{Spectra}~(c)], one obtains easily that $\Delta\eta(\tau)=2a\tanh[\tau/(2t^*)]$. For a fully vertical front, $t^*$ tends towards zero leading to $\Delta\eta$ independent of $\tau$. For finite steepness shock waves, Eq.~\eqref{SclaingBurgers} is thus modified as 
\begin{align}    
  \mathcal{S}^{\mathrm{mB}}_p(\tau)&=\Gamma\tau^1 \langle|\Delta\eta(\tau)|^p\rangle_t \ , \ \ \ \ \mathrm{with} \label{Spth} \\   
  \langle|\Delta\eta(\tau)|^p\rangle_t &= \left|2a\tanh\left(\frac{\tau-\tau_0}{2t^*}\right)+\langle|\Delta\eta_0|^p\rangle_t^{1/p}\right|^p . \label{Spthamp}
\end{align} 
As this prediction is only valid when $\tau > \tau_s$, the shortest possible time lag $\tau_0$ (2~ms) and the corresponding shortest increment amplitude $\langle|\Delta\eta_0|^p\rangle_t=\langle|\Delta\eta(\tau_0)|^p\rangle_t$ are needed. The theoretical values $\zeta_p^{\mathrm{th}}(\tau)$ thus reads $\zeta_p^{\mathrm{th}}(\tau)\equiv d\log(S_p^{\mathrm{th}})/d\log(\tau)$ where $S_p^{\mathrm{th}}=S_p^{\mathrm{ni}}+S_p^{\mathrm{mB}}$ with $S_p^{\mathrm{ni}}=[\mathcal{S}_2(0)\tau^{\frac{n-1}{2}}]^p$ the nonintermittent part, and $S_p^{\mathrm{mB}}$ the shock-dominated part of the structure functions from Eq.~\eqref{Spth}. $S_p^{\mathrm{th}}$ is then reported in Fig.~\ref{Zetap} (solid line) and is in good agreement with experimental data (bullets) with no fitting parameter once the shock typical geometry ($t^*$, $a$, and $n$) are known. The time-averaged amplitude of the shock increments, $\langle|\Delta\eta(\tau)|^p\rangle_t$ of Eq.~\eqref{Spthamp} is also successfully compared with the experiments in Fig.~\ref{S4_SpNorma}(c) for different $p$ {leading to mean values of $a=3.1$~mm and $t^*=7.7$~ms, close to the ones found directly fitting the shock-wave profile as in Fig.~\ref{Spectra}(b)}. Finally, the experimental and theoretical structure functions of order $p$ are compared by plotting $\mathcal{S}_p/\mathcal{S}_p^{\mathrm{th}}$ in Fig.~\ref{S4_SpNorma}(d). Curves collapse well towards a constant value of the order of unity within the inertial range, thus validating the modified Burgers model, experimentally. {The influence of the distribution widths of $a$ and $t^*$ on the compensated structure function, $\mathcal{S}_4/\mathcal{S}_4^{\mathrm{th}}$, is shown in the Appendix~\ref{model_limits}. $\mathcal{S}_4/\mathcal{S}_4^{\mathrm{th}}$ shows some fluctuations, but less than one order of magnitude.}

\textit{Conclusion.---}
We have reported the experimental observation of intermittency in a regime dominated by random shock waves on the surface of a fluid. Their energy spectrum is well described by the theoretical Kuznetsov spectrum involving a random set of singularities~\cite{KuznetsovJETP2004}. We have shown that these shock waves lead to small-scale intermittency, quantified by the PDFs of the increment of wave-amplitude fluctuations and by high-order statistics (structure functions). The reported intermittency is found to be much more intense than in 3D hydrodynamics turbulence or wave turbulence. We have developed a Burgerslike intermittency model, modified to take account of the experimental finite steepness of the shock waves, which is found to be in good agreement with data. 

Our results could be applied to other turbulent systems. Indeed, better understanding the role of coherent structures in forming a turbulent spectrum and causing intermittent behavior is crucial, particularly in wave turbulence and 3D turbulence. As intermittency is associated with the singularities of the turbulent flow~\cite{LassigPRL2000}, vortex filaments, for instance, could play the role of the Burgers shocks~\cite{Frisch1995} although they have much more complicated statistics (multifractal instead of bifractal scaling). Since Burgers equation has a number of further applications from condensed matter to cosmology~\cite{Bec2007}, and is formally equivalent to the Kardar-Parisi-Zhang equation describing interface growth dynamics in a random medium~\cite{KPZ1986}, to which extent our results can be applied to this range of fields is an open question. {Finally, dissipative effects could be tested in the future (as the fluid viscosity in the Burgers equation impacts the shock-wave front $t^*$) using ferrofluids of different viscosities and constant magnetic properties.} 

\begin{acknowledgments}
This work was supported by the Simons Foundation MPS No.~651463--Wave Turbulence (USA) and the French National Research Agency (ANR Sogood Project No.~ANR-21-CE30-0061-04).
\end{acknowledgments}

\appendix

\section{Numerical simulations of the 1D viscous Burgers equation}\label{BurgersNum}
\begin{figure}[h!]
    \centering
    \includegraphics[width=0.9\linewidth]{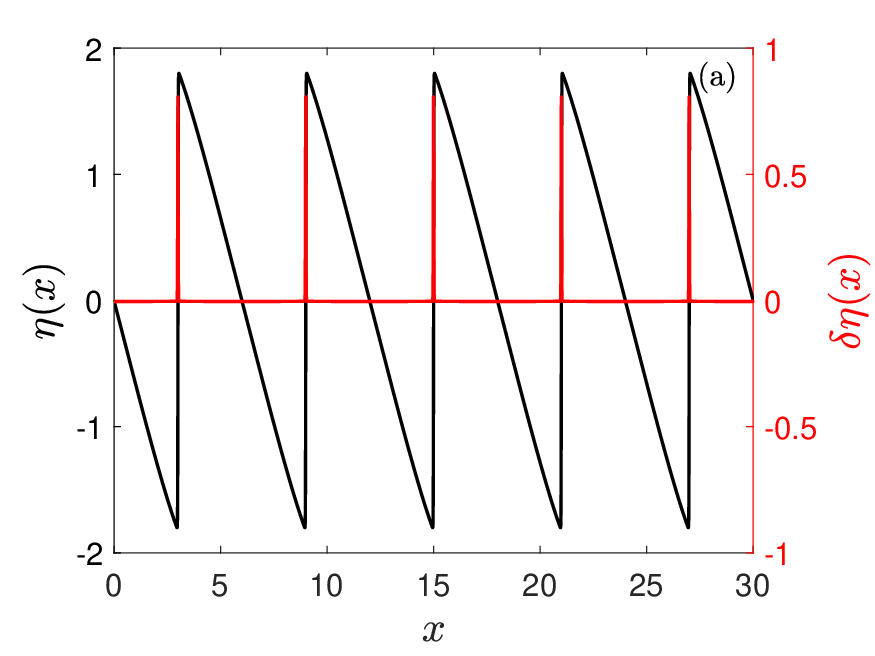}\\ 
    \includegraphics[width=0.9\linewidth]{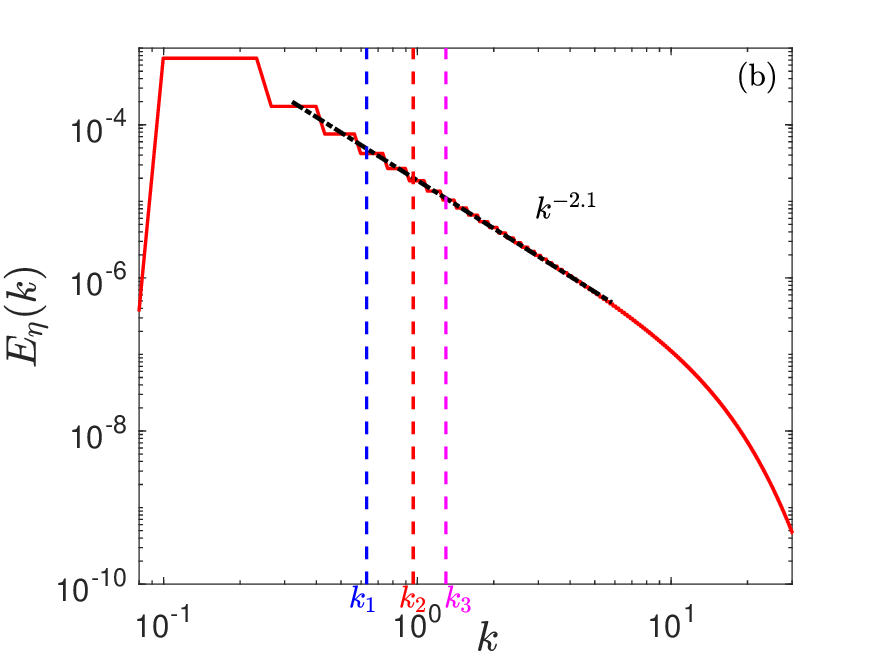} 
    \caption{{(a)} Numerical solution of Eq.~\eqref{Burg_eq} at large $t$, showing $\eta(x)$ (black line) with five successive shock waves and their first-order difference $\delta\eta(x)$ (red line). $A=25$ and $\nu=2.86\ 10^{-6}$. {(b) Corresponding power spectrum $E_\eta(k)$. Black dash-dotted line: best fit in $k^{-2.1}$. Vertical dashed lines represent the values of $k_i=1/(2r_i)$ with $r_i$ chosen to be in the inertial range of the spectrum. The values of $r_i$ are equivalent (changing $k$ to $\omega$) to the values of $\tau_i$ used in Fig.~\ref{S4_SpNorma}(a).}}
    \label{Burgers_shocks}
\end{figure}

The 1D viscous Burgers equation reads~\cite{Burgers1948}
\begin{equation}
    \frac{\partial\eta}{\partial t}+A\eta\frac{\partial\eta}{\partial x}=\nu\frac{\partial^2\eta}{\partial x^2},
    \label{Burg_eq}
\end{equation}
with $A$ an arbitrary constant that ensures dimensional homogeneity and $\nu$ the kinematic viscosity. To numerically solve Eq.~\eqref{Burg_eq}, we use an implicit scheme with the Crank-Nicolson formulation and a Thomas algorithm with the initial condition $\eta(x,t=0)=\sin(x)$~\cite{Ricard2023}. The numerical grid is resolved with 1024 points. Five successive shocks (black) and their first-order difference $\delta\eta(x)=\eta(x+dx)-\eta(x)$ (red) are displayed in Fig.~\ref{Burgers_shocks}{(a)} at large $t$. Shock waves with a vertical front are visible, and their difference is a $\delta$-Dirac distribution. {The corresponding power spectrum $E_\eta(k)$, displayed in Fig.~\ref{Burgers_shocks}(b), scales as $k^{-2}$ as expected.} In the inertial range, the amplitude of the corresponding increments $\delta\eta(x,r)=\eta(x+r)-\eta(x)$ is independent of the separation $r$, conversely to its width [see Fig.~\ref{S4_SpNorma}(a) -- the time lag $\tau$ playing the role of $r$ since temporal signals are involved experimentally]. The shocks observed experimentally differ from the numerical ones of Fig.~\ref{Burgers_shocks}(a). The experimental fronts are not fully vertical, but have a finite steepness because of residual dispersive effects [see Fig.~\ref{Spectra}(b)]. This leads to a  $\delta$-Dirac distribution in the second-order difference $\delta^{(2)}\eta(t)=\eta(t+2dt)-2\eta(t+dt)+\eta(t)$ and not in the first-order difference. They are also found to appear randomly in the experimental signal~\cite{Ricard2023}.

{
\section{Shock-wave propagation}\label{shock_pulse_fomation}
The experimental propagation of the surface elevation in response to a single pulse forcing is shown in Fig.~\ref{Signal_pulse}. This signal is obtained using a spatiotemporal measurement~\cite{Ricard2023}. A shock wave involving a discontinuity is observed traveling along the canal, keeping a self-similar shape. 

\begin{figure}[h!]
    \centering
    \includegraphics[width=1\linewidth]{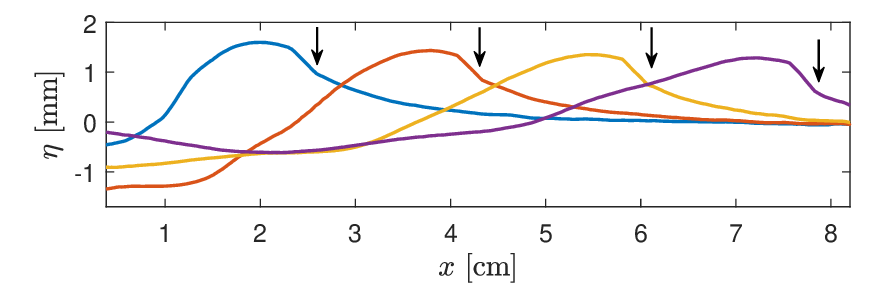} 
    \caption{{Experiment: Spatial evolution of the surface elevation in response to a single pulse forcing for increasing times (spaced from 25 ms, from blue to purple, i.e., from left to right). The arrows indicate the discontinuity location over time.}}
    \label{Signal_pulse}
\end{figure}
}

\section{Correlation time estimation}\label{corr_time}
\begin{figure}[h!]
    \centering
    \includegraphics[width=0.9\linewidth]{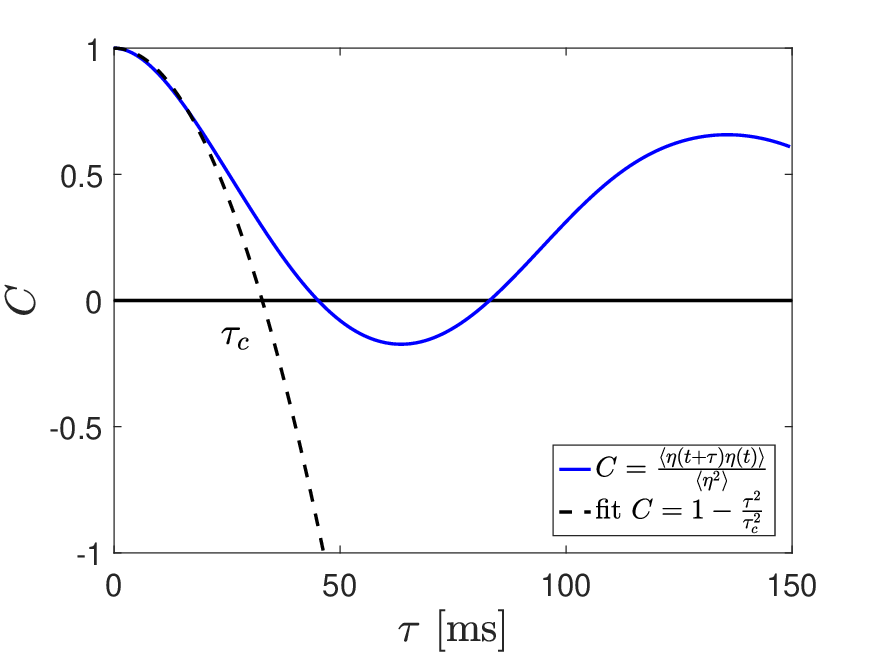} 
    \caption{Normalized autocorrelation function $C(\tau)$ of Eq.~\eqref{corre_eq} versus the time lag $\tau$ (solid blue line) and its parabolic fit from Eq.~\eqref{para_fit} for small $\tau$ (black dashed line). $\tau_c\approx 33$~ms.}
    \label{corre}
\end{figure}

The normalized autocorrelation function $C(\tau)$ of a temporal signal $\eta(t)$ is defined as
\begin{equation}
    C(\tau)=\frac{\langle\eta(t+\tau)\eta(t)\rangle_t}{\langle\eta(t)^2\rangle_t}.
    \label{corre_eq}
\end{equation}
The correlation time $\tau_c$ can be inferred from $C(\tau_c)=0$. For small values of $\tau$, $C$ can be approximated by a parabolic function that provides a better estimation of the correlation time, $\tau_c$, as
\begin{equation}
    C(\tau)\approx1-\frac{\tau^2}{\tau_c^2}.
    \label{para_fit}
\end{equation}
If two points of the signal $\eta(t)$ are separated by a time lag $\tau \gg \tau_c$ they are fully uncorrelated and are thus independent. The experimental correlation function is displayed in Fig.~\ref{corre}. The parabolic fit of Eq.~\eqref{para_fit} provides an estimation of the correlation time $\tau_c\approx 33$~ms. The range of $\tau$ used in the main paper corresponds to $\tau\ll\tau_c$.

\section{Flatness and hyperflatness coefficients}\label{Flatness}
\begin{figure}[h!]
    \centering
    \includegraphics[width=0.9\linewidth]{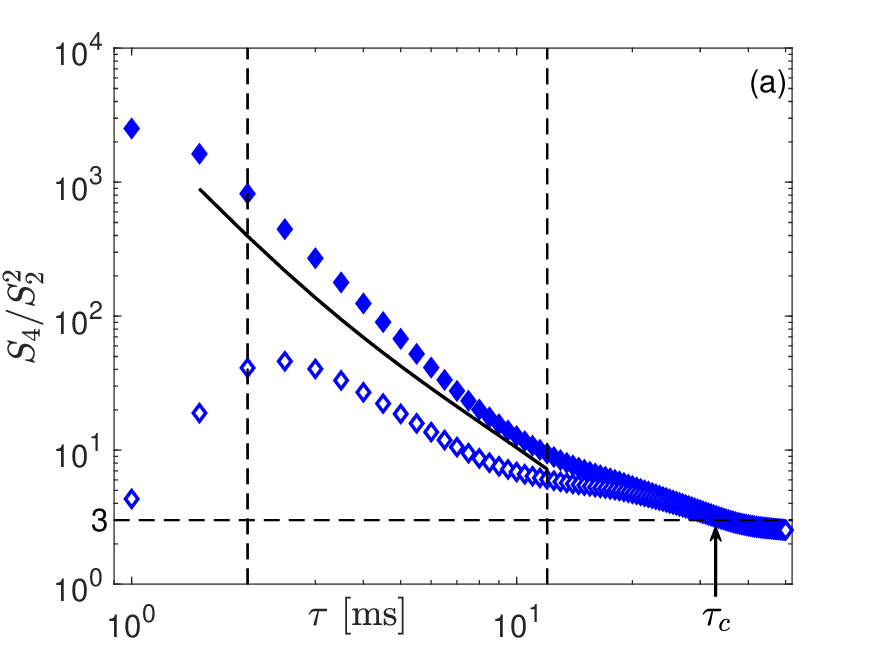}\\ 
    \includegraphics[width=0.9\linewidth]{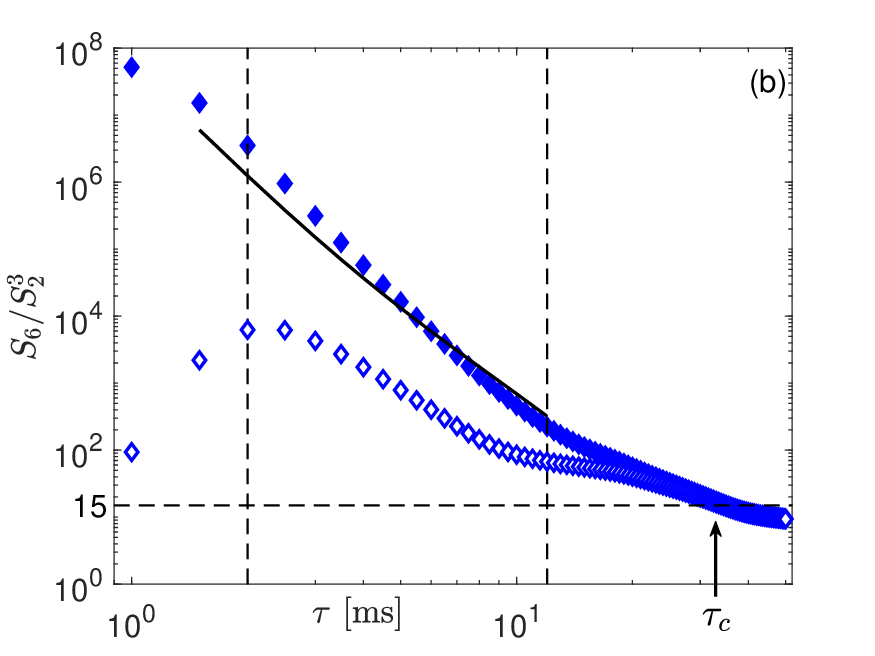} 
    \caption{Coefficients of (a) flatness, $\mathcal{S}_4/\mathcal{S}^2_2$, and (b) hyperflatness, $\mathcal{S}_6/\mathcal{S}^3_2$, as a function of the time lag $\tau$ with ($\blacklozenge$) and without ($\lozenge$) shock waves. Horizontal dashed lines: expected values for a Gaussian distribution {(3 and 15, respectively)}. Solid lines: prediction of the Burgerslike intermittency model, $S^\mathrm{th}_p/S{^\mathrm{th}_2}^{p/2}$, using Eq.~\eqref{Spth}. Vertical dashed lines indicate the inertial range. $\tau_c\approx 33$~ms (see Appendix~\ref{corr_time}).}
    \label{Flat}
\end{figure}
The small-scale intermittency (i.e., the PDF shape deformations over the scales $\tau$) can be quantified by the dependence of the flatness coefficient, $\mathcal{S}_4/\mathcal{S}^2_2$, with $\tau$ as shown in Fig.~\ref{Flat}(a). At large $\tau$, the flatness is close to 3 (the value for a Gaussian) and increases up to $10^3$ at small $\tau$, corresponding to a much flatter PDF [see Fig.~\ref{Spectra}(d)]. Same is performed in Fig.~\ref{Flat}(b) for the hyperflatness, $\mathcal{S}_6/\mathcal{S}^3_2$, ranging from 15 (the value for a Gaussian) up to $10^6$. These experimental dependencies of the flatness and hyperflatness coefficients are in agreement with the theoretical predictions, $S^\mathrm{th}_p/S{^\mathrm{th}_2}^{p/2}$, where $S_p^{\mathrm{th}}=S_p^{\mathrm{ni}}+S_p^{\mathrm{mB}}$ with $S_p^{\mathrm{ni}}$ the nonintermittent part (valid for small $p$) and  $S_p^{\mathrm{mB}}$ from Eq.~\eqref{Spth} of the Burgerslike intermittency model (valid for large $p$) and corresponding to the shock-dominated part of the structure functions. When removing the shock waves ($\lozenge$ in Fig.~\ref{Flat}), the flatness drops by a factor of 20 and the hyperflatness by a factor of 500, confirming the strong intermittency is well ascribed to the shock waves. 

{
\begin{figure}[h!]
    \centering
    \includegraphics[width=0.9\linewidth]{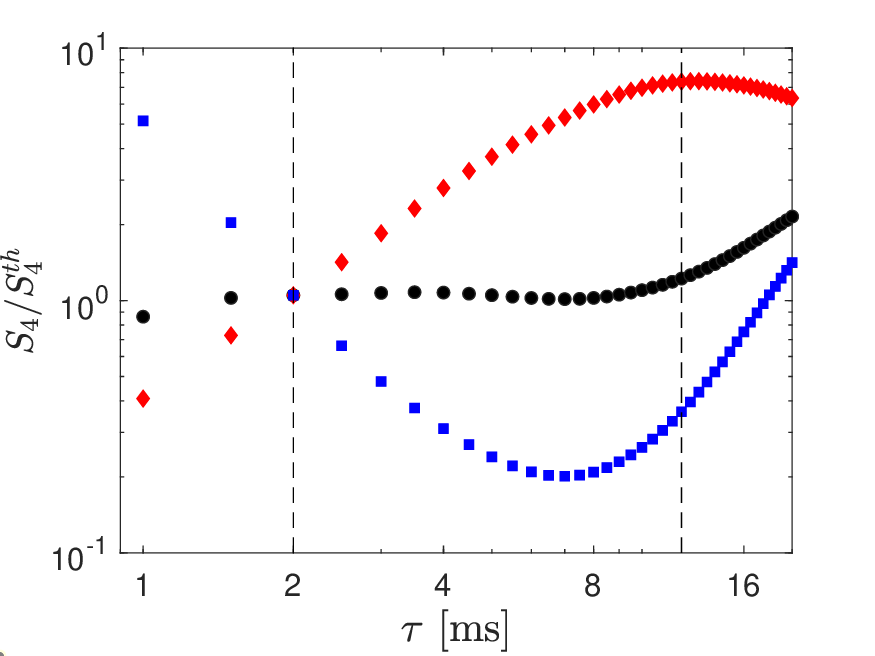}
    \caption{{Compensated structure functions $\mathcal{S}_4/\mathcal{S}_4^{th}$ using best values $a=3.1$~mm and $t^*=7.7$~ms (black $\bullet$) and rms values of the distributions: $a=1$~mm and $t^*=7.7$~ms (red $\blacklozenge$) and $a=3.1$~mm and $t^*=3.5$~ms (blue $\blacksquare$). Vertical dashed lines indicate the inertial range.}}
    \label{S4th_min_max}
\end{figure}
\section{Influence of the shock-wave parameters on the modified Burgers model}\label{model_limits}
The compensated structure function $\mathcal{S}_4/\mathcal{S}_4^{th}$ is plotted for three different values of the shock-wave parameters, $a$ and $t^*$, to evidence the limits of the modified Burgers model. When using the rms values of the distributions of $a$ and $t^*$, $\mathcal{S}_4/\mathcal{S}_4^{th}$ shows some fluctuations, but less than one order of magnitude, thus validating the model.

}


\newpage
\begin{thebibliography}{99}

\bibitem{Batchelor1949}G. K. Batchelor and A. A. Townsend, The nature of turbulent motion at large wave-numbers, \href{https://doi.org/10.1098/rspa.1949.0136}{Proc. R. Soc {\bf 199}, 238 (1949)}.
\bibitem{Frisch1995} U. Frisch, \textit{Turbulence: the legacy of A. N. Kolmogorov}, (Cambridge University, Cambridge, 1995), and references therein.
\bibitem{Kolmogorov41}A. N. Kolmogorov, The local structure of turbulence in incompressible viscous fluid for very large Reynolds numbers, Dokl. Akad. Nauk SSSR {\bf 30}, 301 (1941) [\href{https://doi.org/10.1098/rspa.1991.0075}{Proc. R. Soc. Lond. A {\bf 434}, 9 (1991)}].
\bibitem{Van1970}C. W. Van Atta and W. Y. Chen, Structure functions of turbulence in the atmospheric boundary layer over the ocean, \href{https://doi.org/10.1017/S002211207000174X}{J. Fluid Mech. {\bf 44}, 145 (1970)}; F. Anselmet, Y Gagne, E. J. Hopfinger, and R. A. Antonia, High-order velocity structure functions in turbulent shear flows, \href{https://doi.org/10.1017/S0022112084000513}{J. Fluid Mech. {\bf 140}, 63 (1984)}; F. Anselmet, R. A. Antonia, L. Danaila, Turbulent flows and intermittency in laboratory experiments, \href{https://doi.org/10.1016/S0032-0633(01)00059-9}{Planetary Space Sci. {\bf 49}, 1177 (2001)} and references therein.
\bibitem{SinhuberPRL2017}M. Sinhuber, G. P. Bewley, and E. Bodenschatz, Dissipative Effects on Inertial-Range Statistics of High Reynolds Numbers, \href{https://doi.org/10.1103/PhysRevLett.119.134502}{Phys. Rev. Lett. {\bf 119}, 134502 (2017)}.
E. Lévêque, N. Mordant, J.-F. Pinton, and A. Arnéodo, Intermittency of Velocity Time Increments in Turbulence, \href{https://doi.org/10.1103/PhysRevLett.95.064501}{Phys. Rev. Lett. {\bf 95}, 064501 (2005)}; L. Chevillard and C. Meneveau, Lagrangian Dynamics and Statistical Geometric Structure of Turbulence, \href{https://doi.org/10.1103/PhysRevLett.97.174501}{Phys. Rev. Lett. {\bf 97}, 174501 (2006)}.

\bibitem{IyerPRL2021}K. P. Iyer, G. P. Bewley, L. Biferale, K. R. Sreenivasan, and P. K. Yeung, Oscillations Modulating Power Law Exponents in Isotropic Turbulence: Comparison of Experiments with Simulations, \href{https://doi.org/10.1103/PhysRevLett.126.254501}{Phys. Rev. Lett. {\bf 126}, 254501 (2021)}.
\bibitem{Ishihara2009}T. Ishihara, T. Gotoh, and Y. Kaneda, Study of high-Reynolds number isotropic turbulence by direct numerical simulation, \href{https://doi.org/10.1146/annurev.fluid.010908.165203}{Annu. Rev. Fluid Mech. {\bf 41}, 165 (2009)} and references therein

\bibitem{Kolmogorov62}A. N. Kolmogorov, A refinement of previous hypotheses concerning the local structure of turbulence in a viscous incompressible fluid at high Reynolds number, \href{https://doi.org/10.1017/S0022112062000518}{J. Fluid Mech. {\bf 13}, 82 (1962)}; A. M. Obukov, Some specific features of atmospheric turbulence, \href{https://doi.org/10.1017/S0022112062000506}{J. Fluid Mech. {\bf 13}, 77 (1962)}.
\bibitem{She1994}Z.-S. She, and E. Leveque, Universal scaling laws in fully developed turbulence, \href{https://doi.org/10.1103/PhysRevLett.72.336}{Phys. Rev. Lett. {\bf 72}, 336 (1994)}; B. Dubrulle, Intermittency in fully developed turbulence: Log-Poisson statistics and generalized scale covariance,  \href{https://doi.org/10.1103/PhysRevLett.73.959}{Phys. Rev. Lett. {\bf 73}, 959 (1994)}.

\bibitem{Frisch1978}U. Frisch, P.-L. Sulem, and M. Nelkin, A simple dynamical model of intermittent fully developed turbulence, \href{https://doi.org/10.1017/S0022112078001846}{J. Fluid Mech. {\bf 87}, 719 (1978)}.

\bibitem{Li2005}Y. Li and C. Meneveau, Origin of non-Gaussian statistics in hydrodynamic turbulence, \href{https://doi.org/10.1103/PhysRevLett.95.164502}{Phys. Rev. Lett. {\bf 95}, 164502 (2005)}; 

\bibitem{Falcon2007b}E. Falcon, S. Fauve, and C. Laroche, Observation of Intermittency in Wave Turbulence, \href{https://doi.org/10.1103/PhysRevLett.98.154501}{Phys. Rev. Lett. {\bf 98}, 154501 (2007)}; S. Lukaschuk, S. Nazarenko, S. McLelland, and P. Denissenko, Gravity wave turbulence in wave tanks: space and time statistics, \href{https://doi.org/10.1103/PhysRevLett.103.044501}{Phys. Rev. Lett. {\bf 103}, 044501 (2009)}; E. Falcon, S. G. Roux, and C. Laroche, On the origin of intermittency in wave turbulence, \href{https://doi.org/10.1209/0295-5075/90/34005}{Europhys. Lett. {\bf 90}, 34005 (2010)}; L. Deike, B. Miquel, P. Guti{\'e}rrez, T. Jamin, B. Semin, M. Berhanu, E. Falcon, and F. Bonnefoy, Role of the basin boundary conditions in gravity wave turbulence, \href{https://doi.org/10.1017/jfm.2015.494}{J. Fluid Mech. {\bf 781}, 196 (2015)}; E. Fadaeiazar, A. Alberello, M. Onorato, J. Leontini, F. Frascoli, T. Waseda, and A. Toffoli, Wave turbulence and intermittency in directional wave fields, \href{https://doi.org/10.1016/j.wavemoti.2018.09.002}{Wave Motion {\bf 83}, 94 (2018)}.
\bibitem{Falcon2022}E. Falcon and N. Mordant, Experiments in Surface Gravity–Capillary Wave Turbulence, \href{https://doi.org/10.1146/annurev-fluid-021021-102043}{Annu. Rev. Fluid Mech. {\bf 51}, 1 (2022)}.

\bibitem{RandouxPRL2014}S. Randoux, P. Walczak, M. Onorato, and P. Suret, Intermittency in Integrable Turbulence,  \href{https://doi.org/10.1103/PhysRevLett.113.113902}{Phys. Rev. Lett. {\bf 113}, 113902 (2014)}.

\bibitem{Alexandrova2007}O. Alexandrova, V. Carbone, P. Veltri, and L. Sorriso-Valvo, Solar wind Cluster observations: Turbulent spectrum and role of Hall effect, \href{https://doi.org/10.1016/j.pss.2007.05.022}{Planetary Space Sci. {\bf 55}, 2224 (2007)}; K. H. Kiyani, S. C. Chapman, Yu. V. Khotyaintsev, M. W. Dunlop, and F. Sahraoui, Global Scale-Invariant Dissipation in Collisionless Plasma Turbulence, \href{https://doi.org/10.1103/PhysRevLett.103.075006}{Phys. Rev. Lett. {\bf 103}, 075006 (2009)}.
\bibitem{DeMichelis2004}P. De Michelis and G. Consolini, Time intermittency and spectral features of the geomagnetic field, \href{https://hdl.handle.net/2122/864}{Ann. Geophys. {\bf 47}, 1713 (2004)}.
\bibitem{MuzyPRL2010}R. Ba\"{\i}le and J.-F. Muzy, Spatial Intermittency of Surface Layer Wind Fluctuations at Mesoscale Range, \href{https://doi.org/10.1103/PhysRevLett.105.254501}{Phys. Rev. Lett. {\bf 105}, 254501 (2010)}.


\bibitem{Roy2023}A. Roy, J. R. Picardo, B. Emerson, T. C. Lieuwen, and R. I. Sujith, Small-scale intermittency of premixed turbulent flames, \href{https://doi.org/10.1017/jfm.2023.63}{J. Fluid Mech. {\bf 957}, A21 (2023)}.
\bibitem{PolancoNature2021}J. I. Polanco, N. P. M{\"u}ller, and G. Krstulovic, Vortex clustering, polarisation and circulation intermittency in classical and quantum turbulence, \href{https://doi.org/10.1038/s41467-021-27382-6}{Nat Commun 12, 7090 (2021)}.
\bibitem{vanBokhovenPoF2009}L. J. A. van Bokhoven, H. J. H. Clercx, G. J. F. van Heijst, and R. R. Trieling, Experiments on rapidly rotating turbulent flows, \href{https://doi.org/10.1063/1.3197876}{Physics of Fluids {\bf 21}, 096601 (2009)}. 
\bibitem{Falcon2004}E. Falcon, B. Castaing, and C. Laroche, ``Turbulent" electrical transport in copper powders, \href{https://doi.org/10.1209/epl/i2003-10071-9}{Europhys. Lett. {\bf 65}, 186 (2004)}.

\bibitem{Burgers1948}J. M. Burgers, A mathematical model illustrating the theory of turbulence, \href{https://doi.org/10.1016/S0065-2156(08)70100-5}{Adv. Appl. Mech. {\bf 1}, 171 (1948)}.
\bibitem{Frisch2001}U. Frisch and J. Bec, \href{https://doi.org/10.1007/3-540-45674-0_7}{Burgulence}, in \textit{New Trends in Turbulence}, edited by M. Lesieur, A. Yaglom, and F. David (Springer, Berlin, 2001), Vol. {\bf 74}, pp. 340-383

\bibitem{Aurell1992}E. Aurell, U. Frisch, J. Lutsko, and M. Vergassola, On the multifractal properties of the energy dissipation derived from turbulence data, \href{https://doi.org/10.1017/S0022112092001782}{J. Fluid Mech. {\bf 238}, 467 (1992)}
\bibitem{Gotoh1994}T. Gotoh, Inertial range statistics of Burgers turbulence \href{https://doi.org/10.1063/1.868388}{Phys. Fluids {\bf 6}, 3985 (1994)}.
\bibitem{Bouchaud1995}J. P. Bouchaud, M. Mézard, and G. Parisi, Scaling and intermittency in Burgers turbulence, \href{https://doi.org/10.1103/PhysRevE.52.3656}{Phys. Rev. E {\bf 52}, 3656 (1995)}.
\bibitem{Mitra2005}D. Mitra, J. Bec, R. Pandit, and U. Frisch, Is multiscaling an artifact in the stochastically forced Burgers equation?, \href{https://doi.org/10.1103/PhysRevLett.94.194501}{Phys. Rev. Lett. {\bf 94}, 194501 (2005)}.
\bibitem{Buzzicotti2016}M. Buzzicotti, L. Biferale, U. Frisch, and S. S. Ray, Intermittency in fractal Fourier hydrodynamics: Lessons from the Burgers equation, \href{https://doi.org/10.1103/PhysRevE.93.033109}{Phys. Rev. E {\bf 93}, 033109 (2016)}.
\bibitem{Murray2018}B. P. Murray, and M. D. Bustamante, Energy flux enhancement, intermittency and turbulence via Fourier triad phase dynamics in the 1-D Burgers equation, \href{https://doi.org/10.1017/jfm.2018.454}{J. Fluid Mech. {\bf 850}, 624 (2018)}.


\bibitem{Ricard2023}G. Ricard and E. Falcon, Transition from wave turbulence to acousticlike shock wave regime, \href{https://doi.org/10.1103/PhysRevFluids.8.014804}{Phys. Rev. Fluids, {\bf 8}, 014804 (2023)}.
\bibitem{Meyrand2015}R. Meyrand, K. H. Kiyani, and S. Galtier, Weak magnetohydrodynamic turbulence and intermittency, \href{https://doi.org/10.1017/jfm.2015.141}{J. Fluid Mech. {\bf 770}, R1 (2015)}; S. Chibbaro and C. Josserand, Elastic wave turbulence and intermittency, \href{https://doi.org/10.1103/PhysRevE.94.011101}{Phys. Rev. E {\bf 94}, 011101(R) (2016)}; N. Mordant and B. Miquel, Intermittency and emergence of coherent structures in wave turbulence of a vibrating plate, \href{https://doi.org/10.1103/PhysRevE.96.042204}{Phys. Rev. E {\bf 96}, 042204 (2017)}; A. T. Skvortsov, C. Kirezci, D. Sgarioto, and A. V. Babanin, Intermittency of gravity wave turbulence on the surface of an infinitely deep fluid: Numerical experiment, \href{https://doi.org/10.1016/j.physleta.2022.128337}{Phys. Lett. A {\bf 449}, 128337 (2022)}.

\bibitem{Connaughton2003}C. Connaughton, S. Nazarenko, and A. C. Newell, Dimensional analysis and weak turbulence, \href{https://doi.org/10.1016/S0167-2789(03)00214-8}{Physica D {\bf 184}, 86 (2003)}.
\bibitem{NazarenkoJFM2010}S. Nazarenko, S. Lukaschuk, S. McLelland, and P. Denissenko, Statistics of surface gravity wave turbulence in the space and time domains, \href{https://doi.org/10.1017/S0022112009991820}{J. Fluid Mech. {\bf 642}, 395 (2010)}.
\bibitem{NazarenkoBook}S. Nazarenko, \textit{Wave Turbulence} (Springer, Berlin, 2011).


\bibitem{Falcon2010b}E. Falcon, S. G. Roux, and B. Audit, Revealing intermittency in experimental data with steep power spectra, \href{https://doi.org/ 10.1209/0295-5075/90/50007}{Europhys. Lett. {\bf 90}, 50007 (2010)}.

\bibitem{Verma2000}M. K. Verma, Intermittency exponents and energy spectrum of the Burgers and KPZ equations with correlated noise, \href{https://doi.org/10.1016/S0378-4371(99)00544-0}{Physica A {\bf 277} 359 (2000)}; S. Alam, P. K. Sahu, and M. K. Verma, Universal functions for Burgers turbulence, \href{https://doi.org/10.1103/PhysRevFluids.7.074605}{Phys. Rev. Fluids {\bf 7}, 074605 (2022)}.

\bibitem{KuznetsovJETP2004}E. A. Kuznetsov, Turbulence spectra generated by singularities, \href{https://doi.org/10.1134/1.1804214}{JETP Lett. {\bf 80}, 83 (2004)}.

\bibitem{LassigPRL2000}M. L{\"a}ssig, Dynamical Anomalies and Intermittency in Burgers Turbulence, \href{https://doi.org/10.1103/PhysRevLett.84.2618}{Phys. Rev. Lett. {\bf 84}, 2618 (2000)}.
\bibitem{Bec2007}J. Bec and K. Khanin, Burgers turbulence, \href{https://doi.org/10.1016/j.physrep.2007.04.002}{Phys. Reports {\bf 447}, 1 (2007)}.
\bibitem{KPZ1986}M. Kardar, G. Parisi, and Y.-C. Zhang, Dynamic Scaling of Growing Interfaces, \href{https://doi.org/10.1103/PhysRevLett.56.889}{Phys. Rev. Lett. {\bf 56}, 889 (1986)}; M. L{\"a}ssig, Quantized Scaling of Growing Surfaces, \href{https://doi.org/10.1103/PhysRevLett.80.2366}{Phys. Rev. Lett. {\bf 80}, 2366 (1998)}.  

\end{thebibliography}
\end{document}